\documentclass[12pt]{article}

\usepackage[top=30truemm,bottom=30truemm,left=25truemm,right=25truemm]{geometry}
\usepackage{amssymb}

\usepackage{amsmath}
\usepackage{graphicx}
\usepackage{theorem}
\usepackage{here}
\usepackage{booktabs}
\usepackage{multirow}

\usepackage{natbib}

\newtheorem{thm}{Theorem}
\newtheorem{lem}{Lemma}

\newtheorem{prp}{Proposition}
\theorembodyfont{\rmfamily}

\newtheorem{remark}{Remark}

\def\al{{\alpha}}
\def\be{{\beta}}

\def\ep{{\varepsilon}}
\def\la{{\lambda}}
\def\si{{\sigma}}
\def\om{{\omega}}
\def\th{{\theta}}

\def\bbe{{\text{\boldmath $\beta$}}}

\def\bta{{\text{\boldmath $\eta$}}}

\def\bep{{\text{\boldmath $\varepsilon$}}}
\def\bla{{\text{\boldmath $\lambda$}}}

\def\bom{{\text{\boldmath $\omega$}}}
\def\bth{{\text{\boldmath $\theta$}}}

\def\bpsi{{\text{\boldmath $\psi$}}}
\def\bphi{{\text{\boldmath $\phi$}}}

\def\sih{{\hat \si}}

\def\lah{{\hat \la}}

\def\phih{{\hat \phi}}

\def\sit{{\tilde \si}}

\def\bbeh{{\widehat \bbe}}

\def\blah{{\widehat \bla}}

\def\bomh{{\widehat \bom}}
\def\bthh{{\widehat \bth}}

\def\bpsih{{\widehat \bpsi}}

\def\bphih{{\widehat \bphi}}

\def\bbet{{\widetilde \bbe}}

\def\De{{\Delta}}
\def\Si{{\Sigma}}

\def\Deh{{\widehat \De}}

\def\bSi{{\text{\boldmath $\Si$}}}

\def\u{{\text{\boldmath $u$}}}
\def\v{{\text{\boldmath $v$}}}

\def\x{{\text{\boldmath $x$}}}
\def\y{{\text{\boldmath $y$}}}

\def\A{{\text{\boldmath $A$}}}
\def\B{{\text{\boldmath $B$}}}
\def\C{{\text{\boldmath $C$}}}

\def\H{{\text{\boldmath $H$}}}
\def\I{{\text{\boldmath $I$}}}
\def\J{{\text{\boldmath $J$}}}

\def\M{{\text{\boldmath $M$}}}

\def\P{{\text{\boldmath $P$}}}

\def\V{{\text{\boldmath $V$}}}
\def\W{{\text{\boldmath $W$}}}
\def\X{{\text{\boldmath $X$}}}

\def\Z{{\text{\boldmath $Z$}}}

\def\fh{{\hat f}}

\def\jh{\hat{\jmath}}

\def\Jc{{\cal J}}

\def\Nc{{\cal N}}

\def\Pc{{\cal P}}

\def\tr{{\rm tr\,}}
\def\diag{{\rm diag\,}}
\def\rank{{\rm rank\,}}

\def\vec{{\bf vec\,}}

\def\[{{\text{\boldmath $[$}}}
\def\]{{\text{\boldmath $]$}}}

\def\zero{{\bf\text{\boldmath $0$}}}
\def\one{{\bf\text{\boldmath $1$}}}
\def\dd{{\rm d}}

\def\|{{\,|\,}}

\def\/{{\Bigr/\!\!}}

\def\1r{{\rm (1)}}
\def\2r{{\rm (2)}}
\def\3r{{\rm (3)}}
\def\4r{{\rm (4)}}
\def\5r{{\rm (5)}}

\def\non{{\nonumber}}

\def\ybt{{\widetilde \y}}

\def\IC{{\rm IC}}
\def\RSS{{\rm RSS}}

\begin{document}
\title{A Variant of AIC based on the Bayesian Marginal Likelihood}

\author{
Yuki Kawakubo\footnote{Graduate School of Social Sciences, Chiba University,
1-33, Yayoi-cho, Inage-ku, Chiba, 263-8522, JAPAN,\quad
(E-mail: \texttt{kawakubo@chiba-u.jp})}
 \ Tatsuya Kubokawa\footnote{Faculty of Economics, University of Tokyo, 7-3-1 Hongo, Bunkyo-ku, Tokyo 113-0033, JAPAN,\quad (E-mail: \texttt{tatsuya@e.u-tokyo.ac.jp})}
 \ and
Muni S. Srivastava\footnote{Department of Statistics, University of Toronto, 100 St George Street, Toronto, Ontario, CANADA M5S 3G3,\quad (E-mail: \texttt{srivasta@utstat.toronto.edu})}
}

\date{}
\maketitle

\begin{abstract}
We propose information criteria that measure the prediction risk of a predictive density based on the Bayesian marginal likelihood from a frequentist point of view.
We derive criteria for selecting variables in linear regression models, assuming a prior distribution of the regression coefficients. Then, we discuss the relationship between the proposed criteria and related criteria.
There are three advantages of our method.
First, this is a compromise between the frequentist and Bayesian standpoints because it evaluates the frequentist's risk of the Bayesian model.
Thus, it is less influenced by a prior misspecification.
Second, the criteria exhibits consistency when selecting the true model.
Third, when a uniform prior is assumed for the regression coefficients, the resulting criterion is equivalent to the residual information criterion (RIC) of \citet{ST02}.

\end{abstract}

\section{Introduction}
\label{sec:int}

The problem of selecting appropriate models has been studied extensively in the literature since the work of \citet{Aka73, Aka74}, who derived the so-called Akaike information criterion (AIC).
There are several approaches to solving the model selection problem: information criteria, such as the AIC or BIC \citep{Sch78}; shrinkage methods, such as the lasso \citep{Tib96}; Bayesian techniques; among others.
With regard to Bayesian techniques, \citet{OS09} provide a good review of the key works in this field, including \citet{KM98}, \citet{DFN97}, and \citet{GM93, GM97}. In addition, a Bayesian lasso procedure based on a spike-and-slab prior has attracted much recent attention \citep{XG15}.
However, although these methods are useful and important, we focus on information criteria in this study.

\medskip
Two of the fundamental information criteria are the AIC and the BIC.
Because the AIC and its variants are based on the risks of predictive densities with respect to the Kullback--Leibler (KL) divergence, they are able to select good models in terms of their predictive ability.
In fact, \citet{Shi81}, \citet{Sha97}, and others have shown that the model selected by the AIC minimizes the prediction error asymptotically.
However, it is known that AIC-type criteria do not have the property of consistency; that is, the probability that the criteria will select the true model does not converge to 1.
On the other hand, the BIC, based on the Bayesian marginal likelihood, does exhibit consistency in certain specific models \citep{Nis84}, but does not select models as effectively in terms of their predictive ability.
Therefore, we propose a hybrid of the AIC and the BIC that uses the empirical Bayesian method, which has the property of consistency, but also selects models well in terms of predictive ability.

\medskip
Our approach is to measure the prediction risk of a predictive density based on the Bayesian marginal likelihood from a frequentist point of view.
Specifically, we focus on the variable selection problem for normal linear regression models, assuming a prior distribution of the regression coefficients in order to derive the criterion.
In Section \ref{sec:app}, we consider two prior distributions, namely the normal distribution and the uniform prior distribution.
There are three advantages of our method.
First, this is a compromise between the frequentist and Bayesian standpoints because it evaluates the frequentist's risk of the Bayesian model.
Thus, the method should be less influenced by a prior misspecification.
Second, the criteria exhibits consistency when selecting the true model.
At the same time, our proposed criteria can select a good model, in the sense that the prediction risk is small, because the criteria are based on the KL divergence.
Lastly, a non-informative improper prior can be also used to construct criteria using our approach.
If we consider the Bayesian risk, which is the KL risk, integrated out with respect to the parameters based on the prior distribution, it diverges when the prior is improper.
On the other hand, when we assume a uniform improper prior for the regression coefficients in the normal linear regression model, we can formally derive the marginal likelihood. In this case, the resulting marginal likelihood is the so-called residual likelihood \citep{PT71}, and the proposed information criterion is equivalent to the residual information criterion (RIC) of \citet{ST02}.
Thus, our approach can be considered a theoretical justification of the RIC.

\medskip
The rest of the paper is organized as follows.
In Section \ref{sec:gen}, we provide a unified framework, which we use to derive the proposed information criteria, that can produce various information criteria, including the AIC, BIC, and RIC.
Then, we propose a new approach that uses the Bayesian marginal likelihood in the general framework, and compare various information criteria that are based on a Bayesian model.
In Section \ref{sec:app}, we derive our information criteria for the variable selection problem in a normal linear regression model, assuming a prior distribution of the regression coefficients. In this section, we
also prove the consistency of the criteria.
In Section \ref{sec:sim}, we use simulations to verify the numerical performance of the proposed criteria.
Lastly, Section \ref{sec:dis} concludes the paper.

\section{General}
\label{sec:gen}

\subsection{Conventional information criteria}
In this section, we describe the concept of information criteria from a general point of view.
Let $\y$ be an $n$-variate observable random vector, with density $f(\y|\bom)$ for a vector of unknown parameters $\bom$.
Let $\fh(\ybt;\y)$ be a predictive density of $f(\ybt|\bom)$, where $\ybt$ is an $n$-variate independent replication of $\y$.
Here, we evaluate the predictive performance of $\fh(\ybt;\y)$ in terms of the following risk:
\begin{equation}
\label{eqn:KLg}
R(\bom;\fh) = \int \left[ \int \log \left\{ {f(\ybt|\bom) \over \fh(\ybt;\y)}\right\} f(\ybt|\bom) \dd\ybt \right] f(\y|\bom)\dd\y.
\end{equation}
Because this is interpreted as a risk with respect to the KL divergence, we call it the KL risk.
The spirit of AIC suggests that we can provide an information criterion for model selection as an (asymptotically) unbiased estimator of the information, as follows:
\begin{equation}
\label{eqn:FBIg}
\begin{split}
I(\bom;\fh) =& \ \iint -2 \log \{ \fh(\ybt;\y) \} f(\ybt|\bom) f(\y|\bom)\dd\ybt\dd\y \\
=& \ E_\bom\left[ -2 \log\{ \fh(\ybt;\y) \} \right],
\end{split}
\end{equation}
which is part of (\ref{eqn:KLg}) (multiplied by 2), where $E_\bom$ denotes the expectation, with respect to the distribution, of $f(\ybt,\y|\bom)=f(\ybt|\bom)f(\y|\bom)$.
Let $\De=I(\bom;\fh) - E_\bom[ -2\log\{ \fh(\y;\y)\} ]$.
Then, the AIC variant based on the predictor $\fh(\ybt;\y)$ is defined by
\begin{equation}
\label{eqn:ICg}
\IC(\fh) = -2\log \{ \fh(\y;\y)\} + {\widehat \De},
\end{equation}
where ${\widehat \De}$ is an (asymptotically) unbiased estimator of $\De$ and $\fh(\y;\y)$ is the value of the function $\fh(\ybt;\y)$ evaluated at $\ybt = \y$.

\medskip
Note that $\IC(\fh)$ produces the AIC and BIC for specific predictive densities.

\medskip
(AIC) \
Use $\fh(\ybt; \y)=f(\ybt|\bomh)$ as the maximum likelihood estimator $\bomh$ of $\bom$.
Then, $\IC(f(\ybt|\bomh))$ is the exact AIC, or is the corrected AIC suggested by \citet{Sug78} and \citet{HT89}, which is approximated by the AIC of \citet{Aka73,Aka74} as $-2\log \{f(\y|\bomh)\} + 2 \dim(\bom)$.

\medskip
(BIC) \
Use $\fh(\ybt; \y)=f_{\pi_0}(\ybt)= \int f(\ybt|\bom) \pi_0(\bom)\dd \bom$ as a proper prior distribution $\pi_0(\bom)$.
It is evident that $I(\bom; f_{\pi_0})=E_\bom[ -2 \log\{ f_{\pi_0}(\y) \} ]$. Thus, we have $\De=0$, so that $\IC(f_{\pi_0})=-2\log\{ f_{\pi_0}(\y) \}$, which is the Bayesian marginal likelihood.
Note that $-2\log\{ f_{\pi_0}(\y)\}$ is approximated by ${\rm BIC} = -2\log \{ f(\y|\bomh) \} + \log (n)\cdot \dim(\bom)$.

\subsection{Proposed approach}
\label{subsec:prop}

The criterion $\IC(\fh)$ in (\ref{eqn:ICg}) can produce not only the conventional AIC and BIC, but also various other criteria.
Hereafter, we consider that $\bom$ is divided as $\bom=(\bbe^t, \bth^t)^t$, for a $p$-dimensional parameter vector of interest $\bbe$, and a $q$-dimensional nuisance parameter vector $\bth$, respectively.
We assume that $\bbe$ has prior density $\pi(\bbe|\bla,\bth)$, with hyperparameter $\bla$.
The model is given as follows:
\begin{align*}
\y | \bbe \sim& \ f(\y|\bbe, \bth),
\\
\bbe \sim& \ \pi(\bbe| \bla, \bth),
\end{align*}
where $\bth$ and $\bla$ are estimated from the data.
An inference based on such a model is called an empirical Bayes procedure.
Here, we consider the predictive density $\fh(\ybt;\y)$ as
$$
\fh(\ybt;\y)=f_{\pi}(\ybt|\blah, \bthh) = \int f(\ybt|\bbe,\bthh)\pi(\bbe|\blah,\bthh)\dd\bbe
$$
for some estimators, $\blah$ and $\bthh$.
Then, the information given in (\ref{eqn:FBIg}) is
\begin{equation}
\label{eqn:FBI}
I(\bom;f_\pi) = \iint -2 \log \{ f_{\pi}(\ybt|\blah,\bthh) \} f(\ybt|\bbe, \bth) f(\y|\bbe, \bth)\dd\ybt\dd\y,
\end{equation}
and the resulting information criterion is
\begin{equation}
\label{eqn:FBIC}
\IC(f_\pi) = -2\log \{ f_\pi(\y|\blah,\bthh) \} +\Deh_{f_\pi},
\end{equation}
where $\Deh_{f_\pi}$ is an (asymptotically) unbiased estimator of $\De_{f_\pi} = I(\bom;f_\pi) - E_\bom[ -2\log \{ f_\pi(\y|\blah,\bthh)\}]$.

\medskip
There are three motivations for considering the information $I(\bom;f_\pi)$ in (\ref{eqn:FBI}) and the information criterion $\IC(f_\pi)$ in (\ref{eqn:FBIC}).

First, the precision of the Bayesian predictor $f_\pi(\ybt|\blah,\bthh)$ is characterized by the risk $R(\bom;f_\pi)$ in (\ref{eqn:KLg}), which is based on a frequentist point of view.
On the other hand, the Bayesian risk is defined by
\begin{equation}
\label{eqn:KLb}
r(\bpsi;\fh) = \int R(\bom;\fh) \pi(\bbe|\bla, \bth)\dd \bbe,
\end{equation}
which measures the prediction error of $\fh(\ybt;\y)$, under the assumption that the prior information is correct, where $\bpsi = (\bla^t,\bth^t)^t$.
The resulting Bayesian criteria, such as the PIC \citep{Kit97} or the DIC \citep{SBCvdL02}, are sensitive to a prior misspecification because they depend on the prior information.
However, because $R(\bom;f_\pi)$ can measure the prediction error of the Bayesian model from a frequentist standpoint, the resulting criterion $\IC(f_\pi)$ is less influenced by a prior misspecification.

\medskip
Second, this criterion has the property of consistency.
In Section \ref{sec:app}, we derive criteria for the variable selection problem in normal linear regression models, and prove that the criteria select the true model with probability tending to one.
The BIC and marginal likelihood are known to exhibit consistency, while most AIC-type criteria are not consistent.
However, AIC-type criteria can choose a good model in the sense of minimizing the prediction error.
Our proposed criterion should include both properties, namely consistency when selecting the parameters of interest $\bbe$, and the tendency to select a good model in terms of its predictive ability.

\medskip
Lastly, we can construct the information criterion $\IC(f_\pi)$ even when the prior distribution of $\bbe$ is improper, because the information $I(\bom;f_\pi)$ in (\ref{eqn:FBI}) can be defined formally for the corresponding improper marginal likelihood.
However, because the Bayesian risk $r(\bpsi;f_\pi)$ does not exist for the improper prior, we cannot obtain the corresponding Bayesian criteria or use the Bayesian risk.
Note that our criterion is equivalent to the residual information criterion (RIC) of \citet{ST02} if we assume a uniform prior on the regression coefficients.
In general, the marginal likelihood based on an improper prior depends on an arbitrary scalar constant, which can be included in the prior, but that might be problematic when selecting the model.
However, the criterion based on our approach, using a uniform prior, can work as a variable selection criterion. This is discussed further in Remark \ref{rem:uniform} in Section \ref{sec:app}.

\subsection{Other information criteria based on a Bayesian model}
To clarify our proposed approach, we first explain related information criteria that are based on a Bayesian model.
When the prior distribution $\pi(\bbe|\bla,\bth)$ is proper, we can treat the Bayesian prediction risk $r(\bpsi;\fh)$ in (\ref{eqn:KLb}).
When $\bpsi = (\bla^t, \bth^t)^t$ is known, the predictive density $\fh(\ybt; \y)$ that minimizes $r(\bpsi;\fh)$ is the Bayesian predictive density (posterior predictive density) $f_\pi^*(\ybt|\y, \bpsi)$, given by
$$
\int f(\ybt|\bbe,\bth) \pi(\bbe|\y,\bla,\bth) \dd\bbe = {\int f(\ybt|\bbe,\bth)f(\y|\bbe,\bth)\pi(\bbe|\bla,\bth)\dd\bbe \over \int f(\y|\bbe,\bth)\pi(\bbe|\bla,\bth)\dd\bbe}.
$$
When $\bpsi$ is unknown, we can consider the Bayesian risk of the plug-in predictive density $f_\pi^*(\ybt|\y, \bpsih)$.
In this case, the resulting criterion is known as the predictive likelihood \citep{Aka80a} or the PIC \citep{Kit97}.
The deviance information criterion (DIC) of \citet{SBCvdL02} and the Bayesian predictive information criterion (BPIC) of \citet{And07} are related criteria based on the Bayesian prediction risk $r(\bpsi;\fh)$.

\medskip
Akaike's Bayesian information criterion (ABIC) \citep{Aka80b} is another information criterion based on the Bayesian marginal likelihood, given by
$$
{\rm ABIC} = -2\log\{ f_\pi (\y|\blah) \} +2\dim(\bla),
$$
where the nuisance parameter $\bth$ is not considered.
The ABIC measures the following KL risk:
\begin{equation*}
\int \left[ \int \log \left\{ {f_\pi(\ybt|\bla) \over f_\pi(\ybt|\blah)} \right\} f_\pi(\ybt|\bla) \dd\ybt \right] f_\pi(\y|\bla) \dd\y,
\end{equation*}
which is not the same as either $R(\bom;\fh)$ or $r(\bpsi;\fh)$.
The ABIC is used to choose the hyperparameter $\bla$ in the same manner as the AIC.
However, note that the ABIC works as a model selection criterion for $\bbe$ because it is based on the Bayesian marginal likelihood.

\section{Application to Linear Regression Models}
\label{sec:app}

\subsection{Criteria}
In this section, we derive variable selection criteria for normal linear regression models.
First, we consider a collection of candidate models, defined as follows.
Let the $n\times p_\om$ matrix $\X_\om$ consist of all explanatory variables, and assume that $\rank(\X_\om) = p_\om$.
In order to define candidate models using the index set $j$, suppose that $j$ denotes a subset of $\om = \{ 1,\dots,p_\om \}$ containing $p_j$ elements ({\it i.e.}, $p_j = \#(j)$) and that $\X_j$ consists of $p_j$ columns of $\X_\om$ indexed by the elements of $j$.
We define the class of candidate models as $\Jc = \Pc(\om)$, namely the power set of $\om$, where $\om$ denotes the full model.
We assume that the true model exists in the class of the candidate models $\Jc$, which is denoted by $j_*$.
Note that the dimension of the true models is $p_{j_*}$, which we abbreviate to $p_*$.

\medskip
The candidate model $j$ is the linear regression model
\begin{equation}
\label{eqn:regmodel}
\y = \X_j\bbe_j + \bep,
\end{equation}
where $\y$ is an $n\times 1$ observation vector of the response variables, $\X_j$ is an $n\times p_j$ matrix of the explanatory variables, $\bbe_j$ is a $p_j\times 1$ vector of the regression coefficients, and $\bep$ is an $n\times 1$ vector of the random errors.
Here, $\bep$ has the distribution $\Nc_n(\zero,\si^2\V)$, where $\si^2$ is an unknown scalar and $\V$ is a known positive definite matrix.

\medskip
We consider the problem of selecting the explanatory variables, and assume that the true model can be expressed by each candidate model.
This is the common assumption used to derive an information criterion.
Under this assumption, the true mean of $\y$ can be written as
\begin{equation*}
E(\y) = \X_j\bbe_j^*,
\end{equation*}
where $\bbe_j^*$ is a $p_j\times 1$ vector, the $p_j-p_*$ components of which are exactly $0$, and the remaining components are not $0$.
Hereafter, we omit the model index, $j$, for notational convenience.
Furthermore, we abbreviate $\bbe_j^*$ as $\bbe$.

\medskip
Now, we construct the variable selection criteria for the regression model (\ref{eqn:regmodel}), which has the form (\ref{eqn:FBIC}).
We consider the following two situations.

\bigskip
\noindent
\textbf{[i] A normal prior for $\bbe$}.
We first assume a normal prior distribution for $\bbe$,
$$
\pi(\bbe|\si^2) \sim \Nc(\zero,\si^2\W),
$$
where $\W$ is a $p\times p$ matrix, suitably chosen with full rank.
Examples of $\W$ are $\W = (\la\X^t\X)^{-1}$ for $\la>0$, when $\V$ is the identity matrix, as introduced by \citet{Zel86}, or more simply, $\W = \la^{-1}\I_p$.
For the moment, we assume that $\la$ is known. We discuss how to determine it in Section \ref{subsec:example}.
Because the likelihood is $f(\y|\bbe,\si^2)\sim \Nc(\X\bbe, \si^2\V)$, the marginal likelihood function is
\begin{align*}
f_\pi(\y|\si^2) =& \ \int f(\y|\bbe,\si^2)\pi(\bbe|\si^2)\dd\bbe \non\\
=& \ (2\pi\si^2)^{-n/2}\cdot |\V|^{-1/2} \cdot |\W\X^t\V^{-1}\X +\I_p|^{-1/2} \cdot \exp \left\{ -\y^t\A\y/(2\si^2) \right\},
\end{align*}
where $\A=\V^{-1} -\V^{-1}\X(\X^t\V^{-1}\X +\W^{-1})^{-1}\X^t\V^{-1}$.
Note that $\A=(\V+\B)^{-1}$ for $\B=\X\W\X^t$; that is $f_\pi(\y|\si^2)\sim \Nc(\zero,\si^2(\V+\B))$.
Then, we take the predictive density as $\fh(\ybt;\y) = f_\pi(\ybt|\sih^2)$, and the information (\ref{eqn:FBI}) can be written as
\begin{equation}
\label{eqn:FBI_pi}
I_{\pi,1}(\bom) = E_\bom \left[ n\log(2\pi\sih^2) +\log|\V| +\log|\W\X^t\V^{-1}\X +\I_p| +\ybt^t\A\ybt/\sih^2 \right],
\end{equation}
where $\sih^2=\y^t\P\y/n$, $\P = \V^{-1} - \V^{-1}\X(\X^t\V^{-1}\X)^{-1}\X^t\V^{-1}$, and $E_\bom$ denotes the expectation with respect to the distribution of $f(\ybt,\y|\bbe,\si^2)=f(\ybt|\bbe,\si^2)f(\y|\bbe,\si^2)$ for $\bom=(\bbe^t,\si^2)^t$.
Note that $\bbe$ is the parameter of interest, and $\si^{2}$ is the nuisance parameter corresponding to $\bth$ in the previous section.
Then, we propose the following information criterion.

\begin{prp}
The information $I_{\pi,1}(\bom)$ in $(\ref{eqn:FBI_pi})$ is unbiasedly estimated by the information criterion
\begin{equation}
\label{eqn:FBIC_pi}
\IC_{\pi,1} = -2\log\{ f_\pi(\y|\sih^2) \} +{2n \over n-p-2},
\end{equation}
where
$$
-2\log\{ f_\pi(\y|\sih^2) \} = n\log(2\pi\sih^2) +\log|\V| +\log|\W\X^t\V^{-1}\X +\I_p| +\y^t\A\y/\sih^2;
$$
that is, $E_\bom(\IC_{\pi,1}) = I_{\pi,1}(\bom)$.
\end{prp}

If $n^{-1}\W^{1/2}\X^t\V^{-1}\X\W^{1/2}$ converges to a $p\times p$ positive definite matrix as $n\rightarrow\infty$, $\log|\W\X^t\V^{-1}\X +\I_p|$ can be approximated as $p\log n$, which is the penalty term of the BIC.
In that case, $\IC_{\pi,1}$ is approximately expressed as
\begin{equation*}
\IC_{\pi,1}^* = n\log(2\pi\sih^2) +\log|\V| + p\log n +2 +\y^t\A\y/\sih^2
\end{equation*}
when $n$ is large.

\medskip
Note that only the first term of $\IC_{\pi,1}$ can work as a variable selection criterion because $f_\pi(\y|\si^2)$ is the Bayesian marginal likelihood.
The difference between them is
$$
{2n \over n-p-2} = 2 + {2(p+2) \over n} + O(n^{-2}).
$$
In other words, $\IC_{\pi,1}$ has a slight additional penalty, of order $n^{-1}$.
We compare the performance of the criteria using simulations in Section \ref{sec:sim}.

\medskip
Alternatively, the KL risk $r(\bpsi;\fh)$ in (\ref{eqn:KLb}) can be used to evaluate the risk of the predictive density $f_{\pi}(\ybt|\sih^{2})$ because the prior distribution is proper.
Then, the resulting criterion is
\begin{equation}
\label{eqn:ABIC}
\IC_{\pi,2} = n\log(2\pi\sih^{2}) +\log|\V| +p\log n +p,
\end{equation}
which is an asymptotically unbiased estimator of $I_{\pi,2}(\si^{2}) = E_\pi[I_{\pi,1}(\bom)]$, where $E_\pi$ denotes the expectation with respect to the prior distribution $\pi(\bbe|\si^{2})$; that is $E_\pi E_{\bom}(\IC_{\pi,2}) \rightarrow I_{\pi,2}(\si^{2})$ as $n\rightarrow \infty$.
Interestingly, $\IC_{\pi,2}$ is analogous to the criterion proposed by \citet{Boz87}, known as the consistent AIC, who suggested replacing the penalty term $2p$ in the AIC with $p +p\log n$.

\bigskip
\noindent
\textbf{[ii] Uniform prior for $\bbe$}.
We next assume a uniform prior for $\bbe$, namely $\bbe \sim uniform(\mathbb{R}^p)$.
Although this is an improper prior distribution, we can obtain the marginal likelihood function formally, as follows:
\begin{align*}
f_r(\y|\si^2) =& \ \int f(\y|\bbe,\si^2)\dd\bbe \non\\
=& \ (2\pi\si^2)^{-(n-p)/2} \cdot |\V|^{-1/2} \cdot |\X^t\V^{-1}\X|^{-1/2} \cdot \exp \left\{ -\y^t\P\y/(2\si^2) \right\},
\end{align*}
which is known as the residual likelihood \citep{PT71}.
Then, we take the predictive density as $\fh(\ybt;\y) = f_r(\ybt|\sit^2)$, and the information (\ref{eqn:FBI}) can be written as
\begin{equation}
\label{eqn:FBI_r}
I_r(\bom) = E_\bom \left[ (n-p)\log (2\pi\sit^2) +\log|\V| +\log|\X^t\V^{-1}\X| +\ybt^t\P\ybt/\sit^2 \right],
\end{equation}
where $\sit^2=\y^t\P\y/(n-p)$, which is the residual maximum likelihood (REML) estimator of $\si^2$, based on the residual likelihood $f_r(\y|\si^2)$.
Next, we propose the information criterion.

\begin{prp}
The information $I_r(\bom)$ in $(\ref{eqn:FBI_r})$ is unbiasedly estimated by the infomation criterion
\begin{equation}
\label{eqn:FBIC_r}
\IC_r = -2\log\{ f_r(\y|\sit^2) \} +{2(n-p) \over n-p-2},
\end{equation}
where
$$
-2\log\{ f_r(\y|\sit^2) \} = (n-p)\log(2\pi\sit^2) +\log|\V| +\log|\X^t\V^{-1}\X| +\y^t\P\y/\sit^2;
$$
that is, $E_\bom(\IC_r) = I_r(\bom)$.
\end{prp}
Note that $\y^t\P\y/\sit^2 = n-p$.
If $n^{-1}\X^t\V^{-1}\X$ converges to a $p\times p$ positive definite matrix as $n\rightarrow \infty$, $\log|\X^t\V^{-1}\X|$ can be approximated by $p\log n$.
Then, we can approximate the criterion as
\begin{equation}
\label{eqn:FBIC_rs}
\IC_r^* = (n-p)\log(2\pi\sit^2) +\log|\V| +p\log n +{(n-p)^2 \over n-p-2},
\end{equation}
for large $n$.
Note that $\IC_r^*$ is equivalent to the RIC proposed by \citet{ST02}.
Noting that $(n-p)^2 / (n-p-2) = (n+2) + \{ 4/(n-p-2) - p \}$, we can see that the difference between $\IC_r^*$ and the RIC is $n+2 - p\log(2\pi\sit^2)$; that is $\IC_r^* = {\rm RIC} + n+2 - p\log(2\pi\sit^2)$.
Note too that the criterion based on $f_{r}(\y|\si^{2})$ and $r(\bpsi;f_r)$ cannot be constructed because its KL risk diverges to infinity.

\begin{remark}
\label{rem:uniform}
As discussed in Section \ref{subsec:prop}, the marginal likelihood based on an improper prior depends on an arbitrary scalar constant, which can, in general, be problematic when selecting the model.
However, our $\IC_r$, or its equivalent RIC, can work as a variable selection criterion.
In order to show that, we compare $\IC_r$ with the AIC and BIC.
When $\V = \I_n$, the AIC and BIC for the normal linear regression model can be expressed as
$$
\IC = n\log(2\pi) + n\log(n^{-1}\RSS) + n + g(p),
$$
\end{remark}
where the first three terms are the likelihood part, and the last term $g(p)$ is the penalty, which depends on $p$. Here, $g(p) = 2(p+1)$ for the AIC and $g(p) = p\log(n)$ for the BIC.
Then, $\RSS$ is the residual sum of squares, defined as $\RSS = \Vert \y - \X\bbeh \Vert^2 = n\sih^2$.
On the other hand, $\IC_r$ in (\ref{eqn:FBIC_r}) can be rewritten as
$$
\IC_r = n\log(2\pi) + n\log(n^{-1}\RSS) + n + h(p),
$$
where the first three terms are the same as those of the AIC and BIC, and $h(p)$ is
\begin{align*}
h(p) =& \ p \{ \log(n-p) - \log(n^{-1}\RSS) - \log(2\pi) - 1 \} + \log|\X^t\X| - p\log(n) \\
&+ n\log\{ n/(n-p) \} + 2 + O(n^{-1}).
\end{align*}
Thus, $h(p)$ can represent the penalty for the large model because $\log|\X^t\X| - p\log(n)$ is asymptotically negligible, and the value in the braces of the first term is positive when $n$ is at least moderately large, noting that $n^{-1}\RSS$ becomes small as $p$ becomes large.

\subsection{Typical examples of the linear regression models}
\label{subsec:example}

In the derivation of the criteria, we assumed that the scaled covariance matrix $\V$ of the vector of error terms is known.
However, it is often the case that $\V$ is unknown, and is some function of the unknown parameter $\bphi$, namely $\V = \V(\bphi)$.
In that case, $\V$ in each criterion is replaced with its plug-in estimator $\V(\bphih)$, where $\bphih$ is some consistent estimator of $\bphi$.
This strategy is also used in many other studies, for example in \citet{ST02}, who proposed the RIC.
We suggest that the $\bphi$ be estimated based on the full model.
The scaled covariance matrix $\W$ of the prior distribution of $\bbe$ is also assumed to be known.
In practice, its structure should be specified, and we have to estimate the parameter $\la$ in $\W$ from the data.
In the same manner as $\V$, $\W$ in each criterion is replaced with $\W(\lah)$.
Note that $\la$ should be estimated based on each candidate model under consideration, because the structure of $\W$ depends on the model.
We propose that $\la$ is estimated by maximizing the marginal likelihood $f_\pi(\y|\sih^2,\la)$, after substituting in the estimate $\sih^2$.

\medskip
Here, we give three examples for the regression model (\ref{eqn:regmodel}): a regression model with constant variance, a variance components model, and a regression model with ARMA errors. The second and the third models include the unknown parameter in the covariance matrix.

\bigskip
\noindent
\textbf{[1] Regression model with constant variance}.
When $\V = \I_n$, (\ref{eqn:regmodel}) represents a multiple regression model with constant variance.
In this model, the scaled covariance matrix $\V$ does not contain any unknown parameters.

\bigskip
\noindent
\textbf{[2] Variance components model}.
Consider a variance components model \citep{Hen50}, described as follows:
\begin{equation}
\label{eqn:vcmodel}
\y = \X\bbe + \Z_2\v_2 + \dots + \Z_r\v_r + \bta,
\end{equation}
where $\Z_i$ is an $n\times m_i$ matrix with $\V_i = \Z_i\Z_i^t$, $\v_i$ is an $m_i \times 1$ random vector with distribution $\Nc_{m_i}(\zero,\th_i\I_{m_i})$ for $i\geq 2$, $\bta$ is an $n\times 1$ random vector with $\bta \sim \Nc_n(\zero,\V_0 +\th_1\V_1)$ for known $n\times n$ matrices $\V_0$ and $\V_1$, and $\bta, \ \v_2,\dots,\v_r$ are mutually independently distributed.
The nested error regression model (NERM) is a special case of a variance components model, given by
\begin{equation}
\label{eqn:NERM}
y_{ik} = \x_{ik}^t\bbe + v_i + \eta_{ik},\quad (i=1,\dots,m; \ k=1,\dots,n_i),
\end{equation}
where $v_i$ and $\eta_{ik}$ are mutually independently distributed as $v_i \sim \Nc(0,\tau^2)$ and $\eta_{ik} \sim \Nc(0,\si^2)$, respectively, and $n = \sum_{i=1}^m n_i$.
Note that the NERM in (\ref{eqn:NERM}) is given by $\th_1 = \si^2,\ \th_2 = \tau^2, \ \V_1 = \I_n$ and $\Z_2 = \diag(\one_{n_1},\dots,\one_{n_m})$, where $\one_l$ is the $l$-dimensional vector of ones, for the variance components model (\ref{eqn:vcmodel}).
This model is often used for clustered data, where $v_i$ is considered the random effect of the cluster \citep{BHF88}.
For such a model, when we are interested in a specific cluster or in predicting the random effects, an appropriate criterion is the conditional AIC, as proposed by \citet{VB05}, which is based on the conditional likelihood given the random effects.
However, when we wish to predict the fixed effects, namely $\x_{ik}^t\bbe$, the NERM can be seen as a linear regression model and the random effects are part of the error term. In other words, we consider $\bep = \Z_2\v_2 + \bta$, $\V = \V(\phi) = \phi\V_2 + \I_n$ for (\ref{eqn:regmodel}), where $\phi = \tau^2/\si^2$ and $\V_2 = \Z_2\Z_2^t = \diag (\J_{n_1},\dots,\J_{n_m})$ for $\J_l = \one_l\one_l^t$.
In this case, our proposed variable selection procedure is useful.

\bigskip
\noindent
\textbf{[3] Regression model with autoregressive moving average errors}.
Consider the regression model (\ref{eqn:regmodel}), assuming the random errors are generated by an ${\rm ARMA}(q,r)$ process defined by
$$
\ep_i - \phi_1\ep_{i-1} - \dots - \phi_q\ep_{i-q} = u_i - \varphi_1u_{i-1} - \dots - \varphi_ru_{i-r},
$$
where $\{u_i\}$ is a sequence of independent normal random variables, with mean $0$ and variance $\tau^2$.
A special case of this model is the regression model with ${\rm AR}(1)$ errors, satisfying $\ep_1 \sim \Nc(0,\tau^2/(1-\phi^2))$, $\ep_i = \phi\ep_{i-1} +u_i$, and $u_i\sim \Nc(0,\tau^2)$ for $i=2,3,\dots,n$.
When we define $\si^2=\tau^2/(1-\phi^2)$, the $(i,j)$-element of the scaled covariance matrix $\V$ in (\ref{eqn:regmodel}) is $\phi^{|i-j|}$.

\subsection{Consistency of the criteria}
In this subsection, we prove that the proposed criteria exhibit consistency.
Our asymptotic framework is that $n$ tends to infinity and the true dimension of the regression coefficients $p_*$ is fixed.
Following \citet{ST02}, we first show that the criteria are consistent for the regression model with constant variance and pre-specified $\W$. Then, we extend the result to the regression model with a general covariance matrix and the case where $\W$ is estimated.

\medskip
We divide $\Jc$ into two subsets, $\Jc_+$ and $\Jc_-$, where $\Jc_+ = \{ j\in \Jc : j_* \subseteq j \}$ and $\Jc_- = \Jc \setminus \Jc_+$.
Note that the true model $j_*$ is the smallest model in $\Jc_+$, and that $E(\y) = \X_{j_*}\bbe_{j_*}$, abbreviated to $\X_*\bbe_*$, where $\bbe_*$ is a $p_*\times 1$ vector of the true regression coefficients.
Let $\jh$ denote the model selected by some criterion.
Following \citet{ST02}, we make the following assumptions:

\medskip
(A1) $E(\ep_1^4) < \infty$.

\smallskip
(A2) $\displaystyle 0<\liminf_{n\rightarrow\infty} \min_{j\in\Jc} |\X_j^t\X_j/n|$ and $\displaystyle \limsup_{n\to \infty}\max_{j\in\Jc}|\X_j^t\X_j/n|<\infty$.

\smallskip
(A3) $\displaystyle \liminf_{n\to \infty}n^{-1}\inf_{j \in \Jc_-}\Vert \X_*\bbe_* -\H_j\X_*\bbe_* \Vert^2>0$, where $\H_j = \X_j(\X_j^t\X_j)^{-1}\X_j^t$.

\medskip
\noindent
We can now obtain the asymptotic properties of the criteria for the regression model with constant variance.

\begin{thm}
\label{thm:ordinary}
If assumptions {\rm (A1)}--{\rm (A3)} are satisfied, $\Jc_+$ is not empty, the $\ep_i$'s are independent and identically distributed (iid), and $\W_j$ in the prior distribution of $\bbe_j$ is pre-specified, then the criteria $\IC_{\pi,1}$, $\IC_{\pi,1}^*$, $\IC_{\pi,2}$, $\IC_r$, and $\IC_r^*$ are consistent; that is $P(\jh=j_*)\to 1$ as $n\to \infty$.
\end{thm}

The proof of Theorem \ref{thm:ordinary} is given in Appendix \ref{sec:proof}.

\medskip
We next consider the regression model with a general covariance structure and the case where $\W_j$ is estimated from the data.
In this case, $\V$ and $\W_j$ are replaced with their plug-in estimators $\V(\bphih)$ and $\W_j(\lah_j)$, respectively.

\begin{thm}
\label{thm:general}
Assume that $\bphih -\bphi_0$ and $\lah_j - \la_{j,0}$ tend to 0 in probability as $n\to \infty$, for all $j\in\Jc$.
In addition, assume that the elements of $\V(\bphi)$ and $\W_j(\la_j)$ are continuous functions of $\bphi$ and $\la_j$, respectively, and that $\V(\bphi)$ and $\W_j(\bla_j)$ are positive definite in the neighborhood of $\bphi_0$ and $\la_{j,0}$, respectively, for all $j\in\Jc$.
If assumptions {\rm (A1)}--{\rm (A3)} are satisfied when $\X_j$ and $\bep$ are replaced with $\V^{-1/2}\X_j$ and $\bep^* = \V^{-1/2}\bep$, respectively, $\Jc_+$ is not empty and $\ep^*_i$ are iid. Then, the criteria $\IC_{\pi,1}$, $\IC_{\pi,1}^*$, $\IC_{\pi,2}$, $\IC_r$, and $\IC_r^*$ are consistent.
\end{thm}

For the proof of Theorem \ref{thm:general}, we use the same techniques as those used in the proof of Theorem \ref{thm:ordinary}.

\section{Simulations}
\label{sec:sim}

In this section, we compare the numerical performance of the proposed criteria, $\IC_{\pi,1}$ and $\IC_r$, with that of conventional criteria, namely the AIC, BIC, DIC, and the marginal likelihood (ML).
We consider two regression models: a regression model with constant variance and a regression model with ${\rm AR}(1)$ errors. These models are taken as examples of the linear model (\ref{eqn:regmodel}) given in Section \ref{subsec:example}.
The matrix of explanatory variables are randomly generated as $\vec(\X_\om) \sim \Nc_{n\times p_\om}(\zero, \I_{p_\om}\otimes \bSi )$ for $\bSi = 0.9\I_n + 0.1\one\one^t$ in each simulation.

\medskip
When deriving the criterion $\IC_{\pi,1}$, we set the prior distribution of $\bbe$ as $\Nc_p(\zero,\si^2\la^{-1}\I_p)$; that is, $\W = \la^{-1}\I_p$.
The unknown parameter $\phi$ in $\V$ for the AR(1) model is estimated using the maximum likelihood estimator based on the full model.
The hyperparameter $\la$ is estimated by maximizing the marginal likelihood $f_\pi(\y|\sih^2,\la)$, after substituting in the estimate $\sih^2 = \y^t\P\y/n$ of $\si^2$.
Note that $\phi$ is estimated based on the full model, while $\si^2$ and $\la$ are estimated from each candidate model using the plugged-in version of $\V(\phih)$.

\medskip
As a competitor for $\IC_{\pi,1}$, we consider the criterion that uses $f_\pi(\y|\sih^2)$ only, which is the so-called the marginal likelihood commonly used in Bayesian analyses.
Another competitor is the DIC, which is also popular in Bayesian analyses.
When deriving the DIC, we consider the same prior distribution of $\bbe$ as that assumed when deriving $\IC_{\pi,1}$, namely $\bbe \sim \Nc_p(\zero,\si^2\la^{-1}\I_p)$.
In fairness to the other criteria, we take $\si^2$ as an unknown parameter and do not assume a prior distribution for the derivation of the DIC.
Let $D(\bbe) = -2\log\{ f(\y|\bbe,\si^2) \}$.
When $\si^2$ is known, the DIC is
$$
{\rm DIC}(\si^2) = 2E_{\bbe|\y}[ D(\bbe) ] - D(\bbet),
$$
where $E_{\bbe|\y}$ denotes the expectation with respect to the conditional distribution of $\bbe$, given $\y$ and $\bbet = E_{\bbe|\y}(\bbe)$.
Because $\bbe|\y \sim \Nc_p( \bbet, \si^2(\X^t\V^{-1}\X + \W^{-1})^{-1} )$ for $\bbet = ( \X^t\V^{-1}\X + \W^{-1} )^{-1}\X^t\V^{-1}\y$, the first term of the DIC is
\begin{align*}
2E_{\bbe|\y}E[D(\bbe)] =& \ 2\tr\big[ \X^t\V^{-1}\X \{ \si^2( \X^t\V^{-1}\X + \W^{-1} )^{-1} + \bbet\bbet^t \} \big] / \si^2 - 4\y^t\V^{-1}\X\bbet / \si^2 \\
& + ( {\rm the} \ {\rm term} \ {\rm which} \ {\rm is} \ {\rm irrelevant} \ {\rm to} \ {\rm the} \ {\rm model} ),
\end{align*}
and the second term is $D(\bbet) = ( \y - \X\bbet )^t\V^{-1}( \y - \X\bbet ) / \si^2 +$ (the term that is irrelevant to the model).
Then, we use DIC($\sih^2$), where $\sih^2$ is substituted into DIC($\si^2$).

\begin{figure}
\caption{The number of simulations that select the true model by the criteria in 1000 realizations of the regression model with constant variance. The left three figures are the result for $p_*=2$, and the right three figures are for $p_*=4$.}
\begin{center}
\includegraphics[height=20cm]{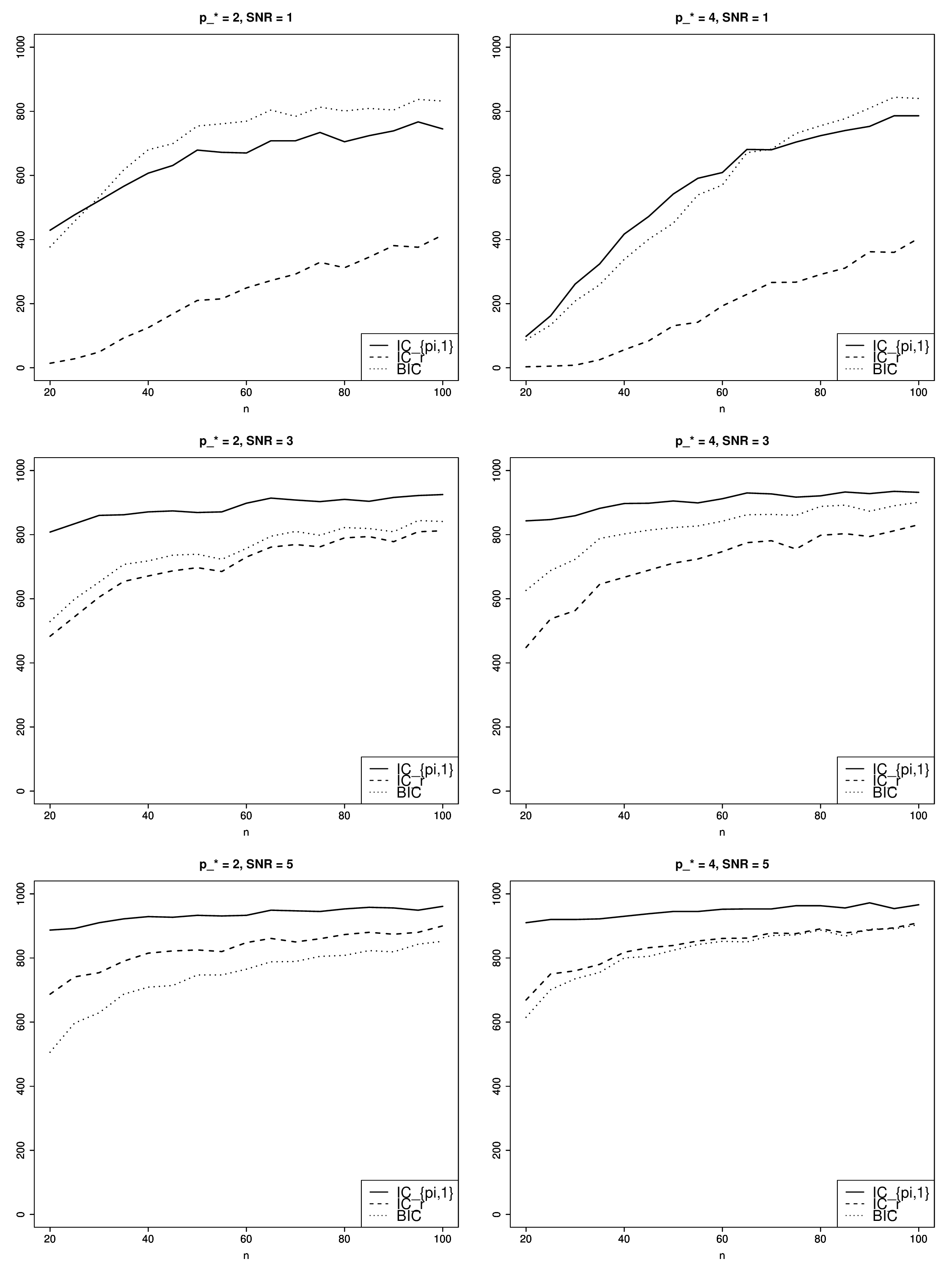}
\end{center}
\label{fig:consistency}
\end{figure}

\medskip
\noindent
\textbf{[Experiment 1]}
First, we confirm that $\IC_{\pi,1}$, $\IC_r$, and the BIC are consistent.
We consider the regression model with constant variance for two cases of the values of regression coefficients $\bbe = (1,1,1,1,0,0,0)^t$ and $\bbe = (1,1,0,0,0,0,0)^t$, namely $p_\om = 7$ and $p_*=2$ or $p_*=4$.
We also control the signal-to-noise ratio (${\rm SNR} = \left\{ {\rm var}(\x_i^t\bbe) / {\rm var}(\ep_i) \right\}^{1/2}$) at 1, 3, and 5.
Note that ${\rm var}(\x_i^t\bbe) = p_*$.
We select a best model by each criterion among all subsets of the full model, that is, we consider $2^7 - 1$ candidate models.
Figure \ref{fig:consistency} shows the number of simulations that select the true model among 1000 simulations.
The results show that each criterion is consistent.
When the data are noisy (i.e., the SNR is weak), $\IC_r$ does not perform as well as $\IC_{\pi,1}$ and the BIC in terms of selecting the true model.
Although we omit the detail, the results for the AR(1) model are similar to those of the model with constant variance.

\begin{table}[htbp]
\centering
\caption{The mean of the prediction error among 1000 simulations for the regression model with constant variance. The full model is $p_\om = 5$.}
\begin{tabular}{cccllllll}
\toprule
& & SNR  & $\IC_{\pi,1}$  & $\IC_r$  & AIC      & BIC      & DIC      & ML     \\
\midrule
\multirow{6}{*}{$p_* = 2$} & \multirow{3}{*}{$n=50$} & 1    & $0.134^{**}$         & 0.167   & 0.154   & $0.134^*$   & 0.185   & 0.135 \\
 & & 3    & $0.0126^{**}$         & 0.0137   & 0.0166   & 0.0134   & 0.0209   & $0.0127^*$ \\
 & & 5    & $0.00430^{**}$        & 0.00447  & 0.00588  & 0.00462  & 0.00765  & $0.00432^*$ \\
\cmidrule{3-9}
 & \multirow{3}{*}{$n=100$} & 1    & $0.0671^*$         & 0.0789   & 0.0790   & $0.0644^{**}$   & 0.0974   & 0.0671 \\
 & & 3    & $0.00628^{**}$        & 0.00658  & 0.00842  & 0.00650  & 0.01084  & $0.00631^*$ \\
 & & 5    & $0.00195^{**}$        & 0.00207  & 0.00301  & 0.00219  & 0.00397  & $0.00197^*$ \\
 \cmidrule{2-9}
 \multirow{6}{*}{$p_*=4$} & \multirow{3}{*}{$n=50$} & 1 & 0.433 & $0.394^{**}$ & 0.409 & 0.469 & $0.394^*$ & 0.430 \\
 &  & 3 & 0.0471 & $0.0434^*$ & $0.0430^{**}$ & 0.0447 & 0.0437 & 0.0470 \\
 &  & 5 & 0.0164 & $0.0155^*$ & $0.0152^{**}$ & 0.0155 & 0.0158 & 0.0164 \\
\cmidrule{3-9}
 & \multirow{3}{*}{$n=100$} & 1 & 0.220 & $0.201^*$ & 0.206 & 0.247 & $0.199^{**}$ & 0.219 \\
 &  & 3 & 0.0231 & $0.0218^*$ & $0.0217^{**}$ & 0.0227 & 0.0225 & 0.0231 \\
 &  & 5 & 0.00764 & $0.00734^*$ & $0.00727^{**}$ & 0.00734 & 0.00774 & 0.00764 \\
\bottomrule
\end{tabular}
\label{tab:PE_const_ps5}
\end{table}

\begin{table}[htbp]
\centering
\caption{The mean of the prediction error among 1000 simulations for the regression model with constant variance. The full model is $p_\om = 10$.}
\begin{tabular}{cccllllll}
\toprule
& & SNR  & $\IC_{\pi,1}$  & $\IC_r$  & AIC      & BIC      & DIC      & ML     \\
\midrule
\multirow{6}{*}{$p_*=2$} & \multirow{3}{*}{$n=50$} & 1 & $0.231^*$ & 0.327 & 0.289 & $0.213^{**}$ & 0.384 & 0.234 \\
 &  & 3 & $0.0185^{**}$ & 0.0224 & 0.0316 & 0.0215 & 0.0437 & $0.0185^*$ \\
 &  & 5 & $0.00591^{**}$ & 0.00672 & 0.0117 & 0.00789 & 0.0162 & $0.00594^*$ \\
\cmidrule{3-9}
 & \multirow{3}{*}{$n=100$} & 1 & $0.106^*$ & 0.139 & 0.140 & $0.0908^{**}$ & 0.192 & 0.106 \\
 &  & 3 & $0.00893^{**}$ & 0.0101 & 0.0162 & 0.00956 & 0.0228 & $0.00895^*$ \\
 &  & 5 & $0.00262^{**}$ & 0.00288 & 0.00542 & 0.00317 & 0.00788 & $0.00264^*$ \\
\cmidrule{2-9}
\multirow{6}{*}{$p_*=8$} & \multirow{3}{*}{$n=50$} & 1 & 1.71 & $1.59^{**}$ & 1.67 & 1.88 & $1.59^*$ & 1.71 \\
 &  & 3 & 0.204 & $0.181^*$ & $0.181^{**}$ & 0.196 & 0.181 & 0.202 \\
 &  & 5 & 0.0690 & 0.0632 & $0.0616^{**}$ & 0.0646 & $0.0629^*$ & 0.0689 \\
\cmidrule{3-9}
 & \multirow{3}{*}{$n=100$} & 1 & 0.912 & $0.798^{**}$ & 0.838 & 1.052 & $0.801^*$ & 0.911 \\
 &  & 3 & 0.0945 & $0.0872^*$ & $0.0866^{**}$ & 0.0930 & 0.0884 & 0.0944 \\
 &  & 5 & 0.0321 & $0.0302^{**}$ & $0.0303^*$ & 0.0306 & 0.0316 & 0.0320 \\
\bottomrule
\end{tabular}
\label{tab:PE_const_ps10}
\end{table}

\begin{table}[htbp]
\centering
\caption{The mean of the prediction error among 1000 simulations for the regression model with constant variance. The full model is $p_\om = 20$.}
\begin{tabular}{cccllllll}
\toprule
& & SNR  & $\IC_{\pi,1}$  & $\IC_r$  & AIC      & BIC      & DIC      & ML     \\
\midrule
\multirow{6}{*}{$p_*=2$} & \multirow{3}{*}{$n=50$} & 1 &$ 0.148^*$ & 0.529 & 0.269 & $0.117^{**}$ & 0.793 & 0.158 \\
 &  & 3 & $0.0123^*$ & 0.0123 & 0.0319 & $0.0122^{**}$ & 0.0894 & 0.0123 \\
 &  & 5 & $0.00363^*$ & $0.00356^{**}$ & 0.01046 & 0.00389 & 0.0320 & 0.00370 \\
\cmidrule{3-9}
 & \multirow{3}{*}{$n=100$} & 1 & $0.0686^*$ & 0.0959 & 0.103 & $0.0533^{**}$ & 0.394 & 0.0703 \\
 &  & 3 & $0.00558^*$ & 0.00550 & 0.01102 & $0.00552^{**}$ & 0.0442 & $0.00558^*$ \\
 &  & 5 & $0.00187^{**}$ & 0.00190 & 0.00410 & 0.00199 & 0.0161 & $0.00188^*$ \\
\cmidrule{2-9}
\multirow{6}{*}{$p_*=18$} & \multirow{3}{*}{$n=50$} & 1 & 7.58 & $7.23^*$ & 7.28 & 11.18 & $7.22^{**}$ & 7.42 \\
 &  & 3 & 0.795 & $0.792^*$ & $0.782^{**}$ & 0.787 & 0.805 & 0.794 \\
 &  & 5 & 0.272 & 0.273 & 0.277 & $0.270^{**}$ & 0.286 & $0.271^*$ \\
\cmidrule{3-9}
 & \multirow{3}{*}{$n=100$} & 1 & 3.61 & 3.59 & $3.53^{**}$ & 6.77 & $3.57^*$ & 3.60 \\
 &  & 3 & $0.384^{**}$ & 0.387 & 0.389 & 0.388 & 0.404 & $0.384^*$ \\
 &  & 5 & $0.135^{**}$ & 0.137 & 0.140 & 0.136 & 0.146 & $0.135^*$ \\
\bottomrule
\end{tabular}
\label{tab:PE_const_ps20}
\end{table}

\begin{table}[htbp]
\centering
\caption{The mean of the prediction error among 1000 simulations for AR(1) with $\phi=0.5$. The full model is $p_\om = 5$.}
\begin{tabular}{cccllllll}
\toprule
& & SNR  & $\IC_{\pi,1}$  & $\IC_r$  & AIC      & BIC      & DIC      & ML     \\
\midrule
\multirow{6}{*}{$p_*=2$} & \multirow{3}{*}{$n=50$} & 1 & 0.203 & $0.194^{**}$ & 0.213 & 0.224 & 0.236 & $0.202^*$ \\
 &  & 3 & 0.0211 & $0.0204^{**}$ & 0.0228 & 0.0220 & 0.0271 & $0.0211^*$ \\
 &  & 5 & 0.00711 & $0.00698^{**}$ & 0.00786 & 0.00740 & 0.00935 & $0.00709^*$ \\
\cmidrule{3-9}
 & \multirow{3}{*}{$n=100$} & 1 & 0.0964 & $0.0915^{**}$ & 0.100 & 0.116 & 0.142 & $0.0963^*$ \\
 &  & 3 & $0.00976^*$ & $0.00952^{**}$ & 0.0105 & 0.0104 & 0.0143 & 0.00977 \\
 &  & 5 & $0.00368^*$ & $0.00362^{**}$ & 0.00395 & 0.00384 & 0.00449 & 0.00368 \\
\cmidrule{2-9}
\multirow{6}{*}{$p_*=4$} & \multirow{3}{*}{$n=50$} & 1 & 0.492 & $0.430^{**}$ & 0.472 & 0.552 & $0.448^*$ & 0.490 \\
 &  & 3 & 0.0530 & $0.0472^*$ & 0.0482 & 0.0526 & $0.0464^{**}$ & 0.0529 \\
 &  & 5 & 0.0191 & 0.0172 & $0.0172^{**}$ & 0.0180 & $0.0172^*$ & 0.0191 \\
\cmidrule{3-9}
 & \multirow{3}{*}{$n=100$} & 1 & 0.239 & $0.206^{**}$ & $0.225^*$ & 0.289 & 0.229 & 0.237 \\
 &  & 3 & 0.0263 & 0.0239 & $0.0238^*$ & 0.0268 & $0.0237^{**}$ & 0.0263 \\
 &  & 5 & 0.00937 & 0.00897 & $0.00875^{**}$ & 0.00905 & $0.00878^*$ & 0.00936 \\
\bottomrule
\end{tabular}
\label{tab:PE_AR_ps5}
\end{table}

\begin{table}[htbp]
\centering
\caption{The mean of the prediction error among 1000 simulations for AR(1) with $\phi=0.5$. The full model is $p_\om = 10$.}
\begin{tabular}{cccllllll}
\toprule
& & SNR  & $\IC_{\pi,1}$  & $\IC_r$  & AIC      & BIC      & DIC      & ML     \\
\midrule
\multirow{6}{*}{$p_*=2$} & \multirow{3}{*}{$n=50$} & 1 & $0.273^*$ & 0.291 & 0.314 & 0.292 & 0.355 & $0.272^{**}$ \\
 &  & 3 & $0.0252^*$ & 0.0254 & 0.0322 & 0.0280 & 0.0397 & $0.0250^{**}$ \\
 &  & 5 & $0.00869^{**}$ & 0.00835 & 0.0118 & 0.0101 & 0.0172 & $0.00871^*$ \\
\cmidrule{3-9}
 & \multirow{3}{*}{$n=100$} & 1 & $0.114^{**}$ & 0.123 & 0.139 & 0.125 & 0.176 & $0.115^*$ \\
 &  & 3 & $0.0107^{**}$ & 0.0110 & 0.0149 & 0.0120 & 0.0198 & $0.0107^*$ \\
 &  & 5 & $0.00349^*$ & $0.00348^{**}$ & 0.00528 & 0.00398 & 0.00702 & 0.00350 \\
\cmidrule{2-9}
\multirow{6}{*}{$p_*=8$} & \multirow{3}{*}{$n=50$} & 1 & 1.53 & $1.41^*$ & 1.48 & 1.66 & $1.43^*$ & 1.53 \\
 &  & 3 & 0.184 & $0.156^*$ & 0.161 & 0.178 & $0.156^{**}$ & 0.183 \\
 &  & 5 & 0.0616 & $0.0535^{**}$ & 0.0553 & 0.0585 & $0.0548^*$ & 0.0614 \\
\cmidrule{3-9}
 & \multirow{3}{*}{$n=100$} & 1 & 0.794 & $0.706^{**}$ & 0.756 & 0.928 & $0.724^*$ & 0.788 \\
 &  & 3 & 0.0830 & $0.0754^{**}$ & $0.0759^*$ & 0.0850 & 0.0761 & 0.0830 \\
 &  & 5 & 0.0287 & $0.0266^*$ & $0.0265^{**}$ & 0.0281 & 0.0268 & 0.0287 \\
\bottomrule
\end{tabular}
\label{tab:PE_AR_ps10}
\end{table}

\begin{table}[htbp]
\centering
\caption{The mean of the prediction error among 1000 simulations for AR(1) with $\phi=0.5$. The full model is $p_\om = 20$.}
\begin{tabular}{cccllllll}
\toprule
& & SNR  & $\IC_{\pi,1}$  & $\IC_r$  & AIC      & BIC      & DIC      & ML     \\
\midrule
\multirow{6}{*}{$p_*=2$} & \multirow{3}{*}{$n=50$} & 1 & $0.205^*$ & 0.341 & 0.355 & $0.182^{**}$ & 0.681 & 0.208 \\
 &  & 3 & $0.0193^{**}$ & 0.0197 & 0.0392 & 0.0200 & 0.0760 & $0.0195^*$ \\
 &  & 5 & $0.00622^*$ & $0.00617^{**}$ & 0.0140 & 0.00676 & 0.0271 & 0.00624 \\
\cmidrule{3-9}
 & \multirow{3}{*}{$n=100$} & 1 & $0.0889^*$ & 0.0900 & 0.120 & $0.0786^{**}$ & 0.303 & 0.0901 \\
 &  & 3 & $0.00861^*$ & $0.00860^{**}$ & 0.0131 & 0.00862 & 0.0340 & 0.00864 \\
 &  & 5 & $0.00293^*$ & $0.00292^{**}$ & 0.00470 & 0.00299 & 0.0122 & 0.00294 \\
\cmidrule{2-9}
\multirow{6}{*}{$p_*=18$} & \multirow{3}{*}{$n=50$} & 1 & 6.02 & 6.02 & $5.94^*$ & 7.91 & 6.02 & $5.92^{**}$ \\
 &  & 3 & 0.667 & 0.669 & 0.668 & $0.658^{**}$ & 0.686 & $0.666^*$ \\
 &  & 5 & $0.227^*$ & 0.229 & 0.234 & 0.229 & 0.241 & $0.227^{**}$ \\
\cmidrule{3-9}
 & \multirow{3}{*}{$n=50$} & 1 & 2.74 & 2.77 & $2.70^{**}$ & 3.65 & 2.76 & $2.73^*$ \\
 &  & 3 & $0.286^{**}$ & 0.289 & 0.292 & 0.287 & 0.302 & $0.286^*$ \\
 &  & 5 & 0.102 & 0.102 & 0.104 & $0.102^{**}$ & 0.108 & $0.102^*$ \\
\bottomrule
\end{tabular}
\label{tab:PE_AR_ps20}
\end{table}

\medskip
\noindent
\textbf{[Experiment 2]}
Next, we investigate the performance of the criteria in terms of the prediction error.
In this experiment, we set $\be_1,\dots,\be_{p_*} \  {\rm i.i.d.} \sim \Nc(0, 1)$ and $(\be_{p_* + 1},\dots,\be_{p_\om})^t = \zero$ in each simulation.
Note that some of the values of $\be_1,\dots,\be_{p_*}$ might be close to 0, which makes us difficult to distinguish the true model from the models that include the true model.
In this case, it is more appropriate to evaluate the performance of the information criteria in terms of prediction error than in terms of selecting the true model.
The prediction error of the selected model is defined as $\Vert \X_{\jh}\bbeh_{\jh} - \X_*\bbe_* \Vert^2 /n$, where $\bbeh_{\jh} = (\X_{\jh}^t\V(\phih)^{-1}\X_{\jh})^{-1}\X_{\jh}^t\V(\phih)^{-1}\y$ (i.e., the GLS estimator).
We consider several settings of $p_\om$, $p_*$,  and SNR for regression model with constant variance and for AR(1) model with $\phi = 0.5$.
For $p_\om = 5, 10$ cases, we consider all subsets of the full model as a class of candidate models, and for $p_\om = 20$ case, we consider a class of nested candidate models $j_\al = \{ 1,\dots,\al \}$ for $\al = 1,\dots,p_\om$.
Tables \ref{tab:PE_const_ps5}--\ref{tab:PE_AR_ps20} show the mean value of the prediction error among 1000 simulations.
In each case, we put two asterisks at the minimum value of the prediction error and one asterisk at the second minimum value.
First, we look at the results for the regression model with constant variance, which are shown in Tables \ref{tab:PE_const_ps5}--\ref{tab:PE_const_ps20}.
When the true model is small (\textit{i.e.}, $p_* = 2$), $\IC_{\pi,1}$ performs best or second best in all cases.
Although the marginal likelihood (ML) performs similar to $\IC_{\pi,1}$, the latter is slightly better.
When the true model is large relative to the full model, $\IC_r$ performs best or second best in many cases.
It is also interesting to point out that $\IC_{\pi,1}$, BIC and ML have good performance when the true model is small while $\IC_r$, AIC and DIC perform well when the true model is large.
This might be because the first three criteria tend to select parsimonious models while the last three criteria tend to select larger models.
The results for AR(1) model, which are shown in Tables \ref{tab:PE_AR_ps5}--\ref{tab:PE_AR_ps20}, are similar to those for the regression model with constant variance except that the performance of $\IC_r$ in the case of $p_\om=5$ and $p_*=2$ (upper part of Table \ref{tab:PE_AR_ps5}) is good while that in the same case for the regression model with constant variance (upper part of Table \ref{tab:PE_const_ps5}) is not very good.

\section{Concluding Remarks}
\label{sec:dis}

We have derived variable selection criteria for normal linear regression models relative to the frequentist KL risk of the predictive density, based on the Bayesian marginal likelihood.
We have proved the consistency of the criteria and, using simulations, have shown that they perform well in terms of the prediction.

\medskip
Although our theoretical approach is general, the derivation of the criterion depends on the normal distribution.
If we assume a conjugate prior distribution for the parameter of interest when deriving the criterion, it is easy to extend our approach to other models.
However, for the class of generalized linear models, which includes the Poisson and the logistic regression models, it is difficult to consider a prior distribution where the marginal likelihood can be evaluated analytically.
In such models, we have to rely on some computational method, which we leave for future research.

\medskip
Variable selection for the mixed effects models, such as the variance components model (\ref{eqn:vcmodel}) in Section \ref{subsec:example}, is another important problem.
As discussed, it is appropriate to consider the KL divergence based on the conditional density, given the random effects, when the objective is to predict the random effects, as in the conditional AIC (cAIC).
An extension of our approach to the cAIC-type criterion is also left to future research.

\medskip
\bigskip
\noindent
{\bf Acknowledgments.}

The authors are grateful to the associate editor and the anonymous referee for their valuable comments and helpful suggestions.
The first and second authors were supported, in part, by Grant-in-Aid for Scientific Research from the Japan Society for the Promotion of Science (JSPS).
The third author was supported, in part, by NSERC of Canada.

\appendix
\section{Derivations of the Criteria}
In this section, we show the derivations of the criteria.
To this end, we first obtain the following lemma, which was shown in Section A.2 of \citet{SK10}.

\begin{lem}
\label{lem:ratio}
Assume that $\C$ is an $n\times n$ symmetric matrix, $\M$ is an idempotent matrix of rank $p$, and that $\u\sim\Nc(\zero,\I_n)$.
Then,
$$
E\left[ {\u^t\C\u \over \u^t(\I_n-\M)\u} \right] = {\tr(\C) \over n-p-2} -{2\tr[\C(\I_n-\M)] \over (n-p)(n-p-2)}.
$$
\end{lem}

\subsection{Derivation of $\IC_{\pi,1}$ in (\ref{eqn:FBIC_pi})}
It is sufficient to show that the bias correction $\De_{\pi,1} = I_{\pi,1}(\bom) - E_\bom [ -2\log\{ f_\pi(\y|\sih^2) \} ]$ is $2n/(n-p-2)$, where $I_{\pi,1}(\bom)$ is given by (\ref{eqn:FBI_pi}).
It follows that
\begin{align*}
\De_{\pi,1} =& \ E_\bom (\ybt^t\A\ybt/\sih^2) - E_\bom (\y^t\A\y/\sih^2) \\
=& \ E_\bom(\ybt^t\A\ybt) \cdot E_\bom(1/\sih^2) -E_\bom(\y^t\A\y/\sih^2).
\end{align*}
First,
\begin{align}
E_\bom(\ybt^t\A\ybt) =& \ E_\bom [(\ybt-\X\bbe+\X\bbe)^t\A(\ybt-\X\bbe+\X\bbe)] \non\\
=& \ \si^2 \tr(\A\V) +\bbe^t\X^t\A\X\bbe. \label{eqn:DeFpi_1}
\end{align}
Second, noting that $n\sih^2 = \y^t\P\y = \si^2\u^t(\I_n-\M)\u$ for
\begin{equation}
\label{eqn:uM}
\begin{split}
\u =& \ \V^{-1/2}(\y-\X\bbe)/\si, \\
\M =& \ \I_n-\V^{-1/2}\X(\X^t\V^{-1}\X)^{-1}\X^t\V^{-1/2},
\end{split}
\end{equation}
and that $\P\X=\zero$, we obtain
\begin{align}
E_\bom(1/\sih^2) =& \ nE_\bom\left( {1 \over \y^t\P\y} \right) = nE_\bom \left[ {1 \over \si^2\u^t(\I_n-\M)\u} \right] \non\\
=& \ {n \over \si^2(n-p-2)}. \label{eqn:DeFpi_2}
\end{align}
Finally,
\begin{align}
E_\bom(\y^t\A\y/\sih^2) =& \ nE_\bom \left( {\y^t\A\y \over \y^t\P\y} \right) = nE_\bom \left[ {\si^2\u^t\V^{1/2}\A\V^{1/2}\u +\bbe^t\X^t\A\X\bbe \over \si^2\u^t(\I_n-\M)\u} \right] \non\\
=& \ n\times \left\{ {\tr(\A\V) \over n-p-2} - {2\tr(\A\V\P\V) \over (n-p)(n-p-2)} +{\bbe^t\X^t\A\X\bbe \over \si^2(n-p-2)} \right\}. \label{eqn:DeFpi_3}
\end{align}
The latter equation derives from Lemma \ref{lem:ratio}.
Combining (\ref{eqn:DeFpi_1}), (\ref{eqn:DeFpi_2}), and (\ref{eqn:DeFpi_3}), we have
\begin{equation*}
\De_{\pi,1} = {2n\cdot \tr(\A\V\P\V) \over (n-p)(n-p-2)}.
\end{equation*}
Therefore,
\begin{align}
\tr(\A\V\P\V) =& \ \tr \{ (\V+\B)^{-1}(\V+\B-\B)\P\V \} \non\\
=& \ \tr(\P\V) - \tr\{ (\V+\B)^{-1}\B\P\V \} \non\\
=& \ \tr(\I_n-\M) = n-p,
\label{eqn:AVPV}
\end{align}
because $\B\P = \X\W\X^t\P = \zero$. Then, we obtain $\De_{\pi,1} = 2n/(n-p-2)$.
\hfill$\Box$

\subsection{Derivation of $\IC_{\pi,2}$ in (\ref{eqn:ABIC})}
From the fact that $E_{\bom}(\IC_{\pi,1}) = I_{\pi,1}(\bom)$ and that $E_\pi E_{\bom}(\IC_{\pi,1}) = E_\pi[I_{\pi,1}(\bom)] = I_{\pi,2}(\si^{2})$, it suffices to show that $E_\pi E_{\bom}(\IC_{\pi,1})$ is approximated by
\begin{align*}
E_\pi E_{\bom}(\IC_{\pi,1}) \approx & \ E_\pi E_{\bom}[ n\log(2\pi\sih^{2}) +\log|\V| +p\log n +2 + \y^{t}\A\y/\sih^{2} ] \\
\approx & \ E_\pi E_{\bom} [ n\log(2\pi\sih^{2}) +\log|\V| +p\log n +p ] +(n+2) = E_\pi E_{\bom}(\IC_{\pi,2}) +(n+2),
\end{align*}
when $n$ is large.
Note that $n+2$ is irrelevant to the model.
It follows that
\begin{align*}
& E_{\bom}\left( {\y^{t}\A\y \over \sih^{2}} \right) \\
=& \ n\times E_{\bom} \left[ {\y^{t}\{ \V^{-1}-\V^{-1}\X(\X^t\V^{-1}\X +\W^{-1})^{-1}\X^t\V^{-1} \} \y \over \y^t\{ \V^{-1} -\V^{-1}\X(\X^t\V^{-1}\X)^{-1}\X^t\V^{-1} \} \y } \right] \\
=& \ n +n\times E_\bom \left[ {\y^t\V^{-1}\X(\X^t\V^{-1}\X +\W^{-1})^{-1}\W^{-1}(\X^t\V^{-1}\X)^{-1}\X^t\V^{-1}\y \over \y^t\{ \V^{-1} -\V^{-1}\X(\X^t\V^{-1}\X)^{-1}\X^t\V^{-1} \} \y } \right] \\
=& \ n +{n \over \si^2(n-p-2)} \times E_\bom \left[ \y^t\V^{-1}\X(\X^t\V^{-1}\X +\W^{-1})^{-1}\W^{-1}(\X^t\V^{-1}\X)^{-1}\X^t\V^{-1}\y \right] \\
=& \ n +{n \over \si^2(n-p-2)}\times \Big[ \si^2\cdot\tr\{ (\X^t\V^{-1}\X +\W^{-1})^{-1}\W^{-1} \} \\
&+\bbe^t\X^t\V^{-1}\X(\X^t\V^{-1}\X +\W^{-1})^{-1}\W^{-1}\bbe \Big],
\end{align*}
and that
\begin{equation*}
E_\pi [ \bbe^t\X^t\V^{-1}\X(\X^t\V^{-1}\X +\W^{-1})^{-1}\W^{-1}\bbe ] = \si^2\cdot \tr[ \X^t\V^{-1}\X(\X^t\V^{-1}\X +\W^{-1})^{-1}].
\end{equation*}
If $n^{-1}\X^t\V^{-1}\X$ converges to a $p\times p$ positive definite matrix as $n\rightarrow \infty$, $\tr[ (\X^t\V^{-1}\X +\W^{-1})^{-1}\W^{-1} ] \rightarrow 0$ and $\tr[ \X^t\V^{-1}\X(\X^t\V^{-1}\X +\W^{-1})^{-1} ] \rightarrow p$.
Then, we have $E_\pi E_\bom(\y^t\A\y/\sih^2 -n) \rightarrow p$, which we want to show.
\hfill$\Box$

\subsection{Derivation of $\IC_r$ in (\ref{eqn:FBIC_r})}
We show that the bias correction $\De_r = I_r(\bom) - E_\bom[-2\log\{ f_r(\y|\sit^2) \}]$ is $2(n-p)/(n-p-2)$, where $I_r(\bom)$ is given by (\ref{eqn:FBI_r}).
Then,
\begin{align*}
\De_r =& \ E_\bom (\ybt^t\P\ybt/\sit^2) - E_\bom(\y^t\P\y/\sit^2) \\
=& \ E_\bom(\ybt^t\P\ybt) \cdot E_\bom(1/\sit^2) -(n-p).
\end{align*}
Since $E_\bom(\ybt^t\P\ybt) = (n-p)\si^2$ and $E_\bom(1/\sit^2)=(n-p)/\{ \si^2(n-p-2) \}$, we have $\De_r = 2(n-p)/(n-p-2)$.
\hfill$\Box$

\section{Proof of Theorem \ref{thm:ordinary}}
\label{sec:proof}
We only prove the consistency of $\IC_{\pi,1}$.
The proof of the consistency of the other criteria can be shown in the same manner.
Because we have
$$
P(\jh = j) \leq P\{ \IC_{\pi,1}(j) < \IC_{\pi,1}(j_*) \}
$$
for any $j\in \Jc \setminus \{ j_* \}$, it suffices to show that $P\{ \IC_{\pi,1}(j) < \IC_{\pi,1}(j_*) \} \to 0$, or equivalently, that $P\{ \IC_{\pi,1}(j) - \IC_{\pi,1}(j_*) > 0 \} \to 1$ as $n\to \infty$ for $j\in \Jc \setminus \{ j_* \}$.
When $\V=\I_n$, we obtain
$$
\IC_{\pi,1}(j) - \IC_{\pi,1}(j_*) = I_1 + I_2 + I_3,
$$
where
\begin{align*}
I_1 =& \ n\log (\sih_j^2 / \sih_*^2) + \y^t\A_j\y / \sih_j^2 - \y^t\A_*\y / \sih_*^2, \\
I_2 =& \ \log|\X_j^t\X_j +\W_j^{-1}| -\log|\X_*^t\X_* +\W_*^{-1}|, \\
I_3 =& \ \log\{ |\W_j|/|\W_*| \} +{2n \over n-p_j-2} -{2n \over n-p_*-2},
\end{align*}
for $\sih^2_j = \y^t(\I_n-\H_j)\y/n$, $\sih_*^2 = \sih^2_{j_*}$, $\A_j = \I_n - \X_j(\X_j^t\X_j +\W_j^{-1})^{-1}\X_j^t$, $\A_{j_*} = \A_*$, and $\W_* = \W_{j_*}$.
We evaluate the asymptotic behaviors of $I_1$, $I_2$, and $I_3$ for $j\in \Jc_-$, and $j\in \Jc_+ \setminus \{ j_0 \}$, separately.

\medskip
[Case of $j\in \Jc_-$].
First, we evaluate $I_1$.
We decompose $I_1 = I_{11} + I_{12}$, where $I_{11} = n\log(\sih^2_j / \sih^2_*)$ and $I_{12} = \y^t\A_j\y/\sih_j^2 - \y^t\A_*\y/\sih_*^2$.
It follows that
\begin{align*}
\sih_j^2 - \sih_*^2 =& \ (\X_*\bbe_* +\bep)^t(\I_n-\H_j)(\X_*\bbe_* +\bep)/n - \bep^t(\I_n-\H_*)\bep/n \\
=& \ \Vert \X_*\bbe_* -\H_j\X_*\bbe_* \Vert^2 / n +o_p(1).
\end{align*}
Then, we have
\begin{equation}
n^{-1}I_{11} = \log \left( 1 + {\sih_j^2 -\sih_*^2 \over \sih_*^2} \right) = \log \left\{ 1 + {\Vert \X_*\bbe_* - \H_j\X_*\bbe_* \Vert^2 \over n\si^2} \right\} +o_p(1), \label{eqn:Pf1_u1}
\end{equation}
and it follows from the assumption (A3) that
\begin{equation}
\liminf_{n\to\infty} \log \left\{ 1 + {\Vert \X_*\bbe_* - \H_j\X_*\bbe_* \Vert^2 \over n\si^2} \right\} >0. \label{eqn:Pf1_u2}
\end{equation}
Because $\y^t\A_j\y / (n\sih_j^2) = 1 + o_p(1)$ and $\y^t\A_*\y / (n\sih_*^2) = 1 +o_p(1)$, we obtain
\begin{equation}
\label{eqn:Pf1_u3}
n^{-1}I_{12} = o_p(1).
\end{equation}
Second, we evaluate $I_2$.
It follows that
\begin{equation*}
\log |\X_j^t\X_j +\W_j^{-1}| = p_j\log n +\log| \X_j^t\X_j/n +\W_j^{-1}/n | = p_j\log n +O(1).
\end{equation*}
In addition, $\log | \X_*^t\X_* +\W_*^{-1} | = p_*\log n +O(1)$.
Then,
\begin{equation}
n^{-1}I_2 = (p_j-p_*)n^{-1}\log n +o(1) = o(1). \label{eqn:Pf1_u4}
\end{equation}
Lastly, it is easy to see that
\begin{equation}
n^{-1}I_3 = o(1). \label{eqn:Pf1_u5}
\end{equation}
From (\ref{eqn:Pf1_u1})--(\ref{eqn:Pf1_u5}), it follows that
\begin{equation}
\label{eqn:Pf1_u}
P\{ \IC_{\pi,1}(j) - \IC_{\pi,1}(j_*) > 0 \} \to 1,
\end{equation}
for all $j\in \Jc-$.

\medskip
[Case of $j\in \Jc_+ \setminus \{ j_* \}$].
First, we evaluate $I_1$.
From
\begin{equation}
\label{eqn:PF1_o0}
\sih_*^2 - \sih_j^2 = \bep^t(\H_j-\H_*)\bep/n = O_p(n^{-1}),
\end{equation}
it follows that
\begin{align}
(\log n)^{-1}I_{11} =& \ (\log n)^{-1} \cdot n \log \left\{ {\sih_*^2-(\sih_*^2-\sih_j^2) \over \sih_*^2} \right\} \non\\
=& \ (\log n)^{-1}\cdot n \cdot \log \{ 1+O_p(n^{-1}) \} = o_p(1). \label{eqn:Pf1_o1}
\end{align}
For $I_{12}$, from (\ref{eqn:PF1_o0}) and $\y^t\A_j\y - \y^t\A_*\y = O_p(1)$, we obtain
\begin{align*}
I_{12} =& \ \y^t\A_j\y/\sih_j^2 - \y^t\A_*\y/\sih_*^2 \\
=& \ (\y^t\A_j\y - \y^t\A_*\y)/\sih_*^2 + O_p(1) = O_p(1).
\end{align*}
Then,
\begin{equation}
\label{eqn:Pf1_o2}
(\log n)^{-1}I_{12} = o_p(1).
\end{equation}
Second, we evaluate $I_2$.
Since $p_j > p_*$ for all $j\in \Jc_+ \setminus \{ j_* \}$,
\begin{equation}
\label{eqn:Pf1_o3}
\liminf_{n\to\infty} (\log n)^{-1}I_2 = p_j - p_* > 0.
\end{equation}
Finally, it is easy to see that
\begin{equation}
\label{eqn:Pf1_o4}
(\log n)^{-1}I_3 = o(1).
\end{equation}
From (\ref{eqn:Pf1_o1})--(\ref{eqn:Pf1_o4}), it follows that
\begin{equation}
\label{eqn:Pf1_o}
P\{ \IC_{\pi,1}(j) - \IC_{\pi,2}(j_*) > 0 \} \to 1,
\end{equation}
for all $j\in \Jc_+ \setminus \{ j_* \}$.

\medskip
Combining (\ref{eqn:Pf1_u}) and (\ref{eqn:Pf1_o}), we obtain
$$
P\{ \IC_{\pi,1}(j) - \IC_{\pi,1}(j_*) > 0 \} \to 1,
$$
for all $j\in \Jc \setminus \{ j_* \}$, which shows that $\IC_{\pi,1}$ is consistent.
\hfill$\Box$

\end{document}